\documentclass[11pt,draftcls,onecolumn]{IEEEtran}
\usepackage{amsfonts}
\usepackage{amsmath,amssymb}
\usepackage{graphicx}
\usepackage{graphics}
\usepackage{subfigure}
\usepackage{multirow}
\usepackage{caption}
\usepackage{makecell}
\usepackage{cite}
\usepackage{color}

\newtheorem{theorem} {Theorem}
\newtheorem{definition} {Definition}
\newtheorem{corollary}  {Corollary}
\newtheorem{lemma}      {Lemma}
\newtheorem{remark}     {Remark}

\begin{document}

\title{Average Consensus by Graph Filtering: New Approach, Explicit Convergence Rate and Optimal Design \thanks{%
This research was supported by the National Science Foundation of China
(grant 61625305, 61701355 and 61471275). }}
\author{Jing-Wen Yi, Li Chai*, and Jingxin Zhang \thanks{%
* Corresponding author. Li Chai is with the Engineering Research Center of
Metallurgical Automation and Measurement Technology, Wuhan University of
Science and Technology, Wuhan, China. e-mail: chaili@wust.edu.cn (L. Chai).}%
\thanks{%
Jing-Wen Yi is with the same University. e-mail: yijingwen@wust.edu.cn (J.
W. Yi).}
\thanks{
Jingxin Zhang is with the School of Software and Electrical Engineering,
Swinburne University of Technology, Melbourne, Australia. e-mail:
jingxinzhang@swin.edu.au (J. Zhang)}}
\maketitle

\begin{abstract}
This paper revisits the problem of multi-agent consensus from a graph signal
processing perspective. Describing a consensus protocol as a graph spectrum filter, we present an effective new approach to the analysis and design of consensus protocols in the graph spectrum domain for the uncertain networks, which are difficult to handle by the existing time-domain methods. This novel approach has led to the following new results in this paper: 1) New necessary and sufficient conditions for both finite-time and asymptotic average consensus of multi-agent systems. 2) Direct link between the consensus convergence rate and the periodic consensus protocols. 3) Conversion of the fast consensus problem to the problem of polynomial design of graph spectrum filter. 4) A Lagrange polynomial interpolation method and a worst-case optimal interpolation method for the design of periodic consensus protocols for the MASs on uncertain  graphs. 5) Explicit formulas for the convergence rate of the designed protocols. Several numerical examples are given to demonstrate the validity, effectiveness and advantages of these results.

\end{abstract}


\IEEEpeerreviewmaketitle

\section{Introduction}

Consensus of multi-agent systems (MASs) is a fundamental problem in collective behaviours of autonomous individuals, which has been extensively studied in the last decades \cite{[01],[02],[03],[04],[05]}. The key problem for consensus is to design appropriate distributed protocols (control sequences) that each agent can only get information from its local neighbors, and the whole network of agents may coordinate to reach an agreement on certain quantities of interest eventually. Many results have been published for MASs, considering different physical phenomenons such as network communication delay \cite{[06],[07]}, switching topology \cite{[9],[12]}, communication channel noise \cite{[08],[09]}, quantized data \cite{[10],[010]} and nonlinear dynamics \cite{[011],[15]}, etc.

As the convergence rate of consensus is crucial to practical applications, many scholars have studied the fast consensus problem of MASs \cite{[16],[17],[18],[19],[20],[21],[22]}.
Xiao and Boyd \cite{[16],[17]} formulated the fastest distributed linear averaging problem and the least-mean-square consensus problem as optimal weight design problems to minimize the asymptotic convergence rate or the total mean-square deviation, and proposed computational methods to solve the corresponding convex optimization problems.
Olfati-Saber \cite{[18]} adopted the `random rewiring' procedure to increase the algebraic connectivity of small-world networks and solved the ultrafast consensus. Aysal et al. \cite{[19]} accelerated the convergence rate of the distributed average consensus by changing the state update to a convex combination of the standard consensus iteration and a linear prediction.
Erseghe et al. \cite{[20]} utilized the alternating direction multipliers method to provide an effective indication on how to choose network matrix for optimized consensus performance.
Kokiopoulou and Frossard \cite{[21]} applied a polynomial filter on the network matrix to shape its spectrum and hence increase convergence rate, and formulated a semidefinite program to optimize the polynomial coefficients.
Montijano et al. \cite{[22]} proposed a fast and stable distributed algorithm based on the Chebyshev polynomials to solve the consensus problem and speed up the convergence rate.

For fixed and known graph topologies, the MASs can reach consensus in finite time under delicately designed control strategies. Sundaram and Hadjicostis \cite{[29]} presented a simple linear iteration to calculate the consensus value, and proposed a finite-time consensus algorithm by introducing the notion of the minimal polynomial of each node. Hendrickx et al. \cite{[30]} utilized the matrix factorization approach to investigate the finite-time consensus, and obtained the algebraic conditions for the minimum polynomial and eigenvalues of
the weight matrix to satisfy the `definitive consensus conjecture'.
Kibangou \cite{[31]} proposed the minimum polynomial approach to design finite-time average consensus protocols, and showed that the smallest possible number of steps to reach consensus is equal to the diameter of the graph. The finite-time consensus problem was also studied in \cite{[23],[25],[26]} using the finite-time Lyapunov stability theory, resulting in some discontinuous or nonlinear consensus protocols.

Graph signal processing (GSP) has recently emerged as a powerful new paradigm  for high-dimensional data analysis and processing \cite{[36],[37],[38],[33],[34],[35]}.
The development of GSP has been to extend the concepts of frequency domain analysis and filtering to graph signals and to develop analysis and design tools for GSP. Some fundamental operations, such as graph Fourier transform, translation, modulation, filtering and convolution, have been defined for graph signals based on the graph topology described by the adjacency or Laplacian matrix, and some effective design methods have been developed \cite{[33],[34],[35],[36],[37],[38],[42]}. Some of these new techniques have been applied to the analysis and design of MAS consensus in the last few years.

Sandryhaila et al. \cite{[43]} designed a matrix polynomial as a graph filter to solve the average consensus in finite time. By solving semidefinite program, some approximation algorithms were obtained to guarantee the finite-time consensus.
Segarra et al. \cite{[44]} proposed an optimal design of graph filters to implement arbitrary linear transformations between graph signals. The developed framework was applied to the finite-time consensus and analog network coding to showcase the relevance of the results for the design of distributed network operators.
Izumi et al. \cite{[45]} showed that the multi-agent consensus corresponds to the low-pass filtering of graph signal, and designed a low-pass filter by polynomial approximation of an exponential function. These results have provided new insights into MASs and opened a new avenue to solving the consensus problem of MASs.

 Despite the excellent works discussed above, there are still some fundamental issues unsolved in MAS consensus. The analysis and design of control protocols are still limited. There are few necessary and sufficient conditions that reveal the essential mechanism of the dynamic evolution of MASs. The convergence rate of consensus protocols are yet to be fully understood. The Lyapunov function based analysis of finite-time consensus can only provide sufficient conditions that can be conservative, and the resulting discontinuous or nonlinear consensus protocols are difficult to realize. The explicit connection between the graph filter and the control protocol has not been studied, and the designs of graph filters for the control protocols are mostly numerical and approximate, lacking analytical solutions.

To address the above issues, in this paper, we study systematically the consensus problem using the recent results of GSP theory, and present the following new results.

1) The new necessary and sufficient conditions for both finite-time and asymptotic average consensus, derived from describing the control sequence as a graph filter $h(\lambda, \cdot)$. These conditions are simple and encompass, as special cases, the MAS consensus under the constant control sequence \cite{[01],[16],[17]}, the MAS consensus under the time-varying control sequence \cite{[9],[012]}, and the finite-time consensus \cite{[30],[31],[43],[45]}.

2) The direct link between the exact convergence rate of consensus, $\rho_M$, and the graph filter of an M-periodic control sequence, $h({\lambda_i}, M)$, in the form $\rho_M =\max_{\{\lambda _{i}\}}\left\vert {h({\lambda_i}, M)}\right\vert$, where $\{\lambda_i\}$ is the Laplacian spectrum of a known graph. This result converts the fast consensus problem to a polynomial design problem of $h({\lambda_i}, M)$, and gives the exact convergence rate instead of the upper bound of convergence rate given in the existing literature.

3) For any uncertain graph $\mathcal{G}$,
with the second smallest eigenvalue $\lambda_2$ and the largest eigenvalue $\lambda_N$ of the graph Laplacian $[\lambda_2, \lambda_N] \subseteq [\alpha, \beta]$, the consensus can be achieved by simply using the control sequence $\varepsilon(jM+k)=\frac{1}{r_k}$, $k=0, \cdots, M-1$, $j \in \mathbb{N}$, with $r_k$ distributed uniformly on $[\alpha, \beta]$. The upper bound of the convergence rate of this protocol is smaller than that of the existing result in \cite{[16]}.

4) The explicit formulas for designing the unique M-periodic control sequence $\{\varepsilon^*(k+jM)\}$ that yields the optimal worst-case convergence rate $\gamma_M^*$ for any uncertain graph $\mathcal{G}$ with $[\lambda_2, \lambda_N] \subseteq [\alpha, \beta]$. The explicit formula for calculating the optimal worst-case convergence rate $\gamma_M^*$ resulting from $\{\varepsilon^*(k+jM)\}$,
and the explicit formula for calculating the worst-case convergence rate under the general time-varying control sequences. All the formulas are analytical and the design and calculations are precise, without iterative approximation.

The rest of this paper is organized as follows. Section II introduces some background about spectral graph theory and graph signal processing. Section III presents necessary and sufficient conditions for average consensus from a graph signal filtering perspective. Upon casting the problem of average consensus to a graph filter design, Section IV investigates
the convergence rate of asymptotic consensus on the known networks under periodic control sequences.
For the uncertain networks, Section V presents a Lagrangian polynomial interpolation method to design the periodic control sequences and compares the convergence performance with the existing result. To pursue the optimal consensus on uncertain networks, Section VI proposes a worst-case optimal interpolation method to obtain the explicit formulas for the design and  convergency rate evaluation of periodic control sequences. The validity and performance of the proposed methods are demonstrated in Section VII by extensive simulation and numerical experiment results. Section VIII concludes the paper with remarks on the presented results and future work.

\section{Preliminaries}

\subsection{Spectral Graph Theory}

Let $\mathcal{G} = (\mathcal{V},\mathcal{E} ,\mathcal{A})$ be a weighted
undirected graph with the set of vertices $%
\mathcal{V} = \{ {{\nu}_1},{{\nu}_2}, \cdots, {{\nu}_N}\} $, the set of edges $%
\mathcal{E} \subseteq \mathcal{V} \times \mathcal{V}$, and a weighted
adjacency matrix $\mathcal{A} = [{a_{ij}}] \in {\mathbb{R} ^{N \times N}}$.
The vertex indices belong to a finite
index set ${\mathcal{I}} = \{ 1,2, \cdots, N\} $. An edge of $%
\mathcal{G}$ is denoted by ${e_{ij}} = ({{\nu}_i},{{\nu}_j}) \in
\mathcal{E}$ if and only if there exist information exchanges between vertex $%
{\nu}_i$ and vertex ${\nu}_j$. The adjacency elements corresponding to the edges of
the graph are positive, i.e., ${e_{ij}} \in \mathcal{E} \Leftrightarrow {%
a_{ij}} = {a_{ji}} > 0$. Assume ${a_{ii}} = 0$ for all $i \in {%
\mathcal{I}}$. The set of neighbors of vertex ${\nu}_i$ is denoted by ${\mathcal{N}%
_i} = \{ {{\nu}_j} \in \mathcal{V}:({{\nu}_i},{{\nu}_j}) \in \mathcal{E} \}$. %
For any ${\nu}_i, {\nu}_j \in \mathcal{V}$, $a_{ij} > 0 \Leftrightarrow j \in {\mathcal{N}_i}$.
The degree of vertex ${\nu}_i$ are represented by $d_i = \sum\nolimits_{j = 1}^N {%
{a_{ij}}}$.
The Laplacian matrix of $\mathcal{G}$ is defined as $\mathcal{L} =%
\mathcal{D} - \mathcal{A}$, where $\mathcal{D}:=diag\{d_1, \cdots, d_N\}$.

\begin{lemma} \cite{[48]}
For an undirected graph $\mathcal{G}= (\mathcal{V},\mathcal{E} ,\mathcal{A})$,
the Laplacian matrix $\mathcal{L}$ satisfies the following properties: \newline
1) $\mathcal{L}$ is symmetric matrix and positive semi-definite. \newline
2) All the eigenvalues of $\mathcal{L}$ are real in an ascending order as
\begin{equation}
0={\lambda _{1}} \leq {\lambda _{2}} \le \cdots \leq {\lambda_{N}} \leq 2 \bar d,
\end{equation}
where $\bar d= \mathop {\max }\limits_i \{ {d_i}\} $ is the maximum degree of the graph. \newline
3) $\mathcal{L}$ has the following SVD decomposition:
\begin{equation}\label{diag}
\mathcal{L}=V\Lambda V^T,
\end{equation}
where $\Lambda=diag\{\lambda_1,\lambda_2,\cdots,\lambda_N\}$, and $V=[\mathrm{v}_1, \mathrm{v}_2, \cdots, \mathrm{v}_N]\in \mathbb{R} ^ {N \times N}$ is an unitary matrix. \newline
4) Zero is a simple eigenvalue of $\mathcal{L}$ if and only if $\mathcal{G}$ is connected,
and the associated eigenvector is ${\mathrm{v}_1} = \frac{1}{\sqrt N }{\vec 1}$, where $\overrightarrow{1}\in \mathbb{R}^N$ is the vector of all ones.
\end{lemma}

\subsection{Graph Signal Processing}

Consider an undirected graph $\mathcal{G} = (\mathcal{V},\mathcal{E} ,\mathcal{A})$ with Laplacian matrix $\mathcal{L}$, the graph signal $x$ is
the collection of the signal values on the vertices of the graph, i.e., $x=[x_1, x_2, \cdots, x_N]^T \in \mathbb{R} ^ N$.
The graph Fourier transform $\hat x$ of $x \in \mathbb{R}^N$
on $\mathcal{G}$ is defined as the expansion of $x$ in terms of
eigenfunctions of the graph Laplacian \cite{[33],[36],[37]}
\begin{equation}  \label{eq01}
{\hat {x}}_{\lambda_i} := \left\langle {x, {\mathrm{v}_i}} \right\rangle =
\sum\limits_{\ell = 1}^N {{x_\ell }{\mathrm{v}_i^{(\ell )}}},
\end{equation}
where $\{\mathrm{v}_i\}_{1,2,\cdots,N}$ are
orthonormal eigenvectors of $\mathcal{L}$. The inverse
graph Fourier transform is then given by
\begin{equation}  \label{eq02}
{x_\ell } = \sum\limits_{i = 1}^N {{\hat {x}}_{\lambda_i}{\mathrm{v}%
_i^{(\ell )}}}.
\end{equation}
Note that $\hat {x}=[\hat{x}_{\lambda_1}, \hat{x}_{\lambda_2}, \cdots, \hat{x}%
_{\lambda_N}]^T = {V^T}x$, where $V$ is the
unitary matrix defined in (\ref{diag}). The inverse graph Fourier transform
is then given by $x={V}{\hat {x}}$. In graph Fourier analysis, the graph
Laplacian eigenvalues correspond to the frequency in the spatial frequency
domain. The eigenvectors associated with smaller eigenvalues (low
frequency) vary slowly across the graph, while those with larger eigenvalues
oscillate more rapidly.

Denote a graph spectral filter ${h}(\cdot)$ as a real-valued function on
the spectrum of graph Laplacian, the graph spectral filtering is defined as \cite{[33]}
\begin{equation}
\hat{y}_{\lambda _{i}} = {h}(\lambda _{i})\hat{x}_{\lambda _{i}}
\label{eq03}
\end{equation}%
and $\hat{y}=[{\hat{y}}%
_{\lambda _{1}},{\hat{y}}_{\lambda _{2}},\cdots ,{\hat{y}}_{\lambda
_{N}}]^{T}$ is the filtered graph signal represented in the frequency
domain.

Taking the inverse graph Fourier transform, we have
\begin{equation}
y=V\left[ {%
\begin{array}{ccc}
{h({\lambda _{1}})} &  &  \\
& \ddots &  \\
&  & {h({\lambda _{N}})}%
\end{array}%
}\right] {V^{T}}x,  \label{eq04}
\end{equation}%
where $y=[{y}_{1},{y}_{2},\cdots ,{y}_{N}]^{T}$ is the filtered
graph signal in spatial domain. When the graph spectral filter in (\ref{eq03})
is an $M$th-order polynomial ${h}(\lambda _{i})=\sum\nolimits_{k=0}^{M}{{%
b_{k}}\lambda _{i}^{k}}$, where $\{b_{k}\}_{k=0,1,\cdots ,M}$ are real coefficients, the filtered signal $y_{\ell }$ at vertex $\ell $ is a linear
combination of the components of the input signal at agents within an $M$-hop
local neighborhood of agent $\ell $ \cite{[33]}
\begin{equation}
y_{\ell } = \sum\limits_{i=1}^{N}{{h}(\lambda _{i})\hat{x}_{\lambda
_{i}}{\mathrm{v}_{i}^{(\ell )}}}
= \sum\limits_{j=1}^{N}{{x_{j}}\sum\limits_{k=0}^{M}{{b_{k}}{{({{\mathcal{L}}^{k}})}%
_{\ell ,j}}}}. \label{eq6}
\end{equation}

\section{Necessary and Sufficient Conditions for Average Consensus}

In this section, we will present necessary and sufficient conditions of average consensus from a graph signal filtering perspective. For each control sequence, we define a corresponding graph filter $h(\lambda, \cdot)$ with $h(0, \cdot)=1$.
We show that average consensus is achieved if and only if the graph filter annihilates all the `high' frequency components at $\lambda_2, \cdots, \lambda_N$,
i.e., $h(\lambda_i, \cdot)=0$, $i=2,\cdots, N$.

Let $\mathcal{G}=(\mathcal{V},\mathcal{E},\mathcal{A})$ be an undirected graph of $N$ nodes with the adjacency matrix $\mathcal{A}=[a_{ij}]_{N\times
N}.$ Suppose each vertex of the graph is an agent described by
\begin{equation}
x_{i}(k+1)=x_{i}(k)+u_{i}(k),i=1,\cdots N,  \label{agent}
\end{equation}%
where $x_{i}(k)\in \mathbb{R} $ is the state and $u_{i}(k)\in \mathbb{R} $ the control input.
We consider a general time-varying control protocol
\begin{equation}
u{_{i}}(k)=\varepsilon (k)\sum\limits_{j\in \mathcal{N}_{i}}a_{ij}(x_{j}(k)-x_{i}(k)),
\label{protocol}
\end{equation}%
where $\varepsilon (k)>0$ is the control gain at time $k$.
Denote $x(k)=[x_1(k), \cdots, x_N(k)]^T \in \mathbb{R}^N$.

\begin{definition}
The average consensus of MAS (\ref{agent}) under the control of protocol (\ref{protocol}%
) is said to be reached asymptotically if
\begin{equation*}
\lim_{k\rightarrow \infty }x_{i}(k)=\frac{1}{N}\sum%
\limits_{j=1}^{N}x_{j}(0):=\bar{x}(0), i=1,2,\cdots ,N
\end{equation*}%
for any initial state $x(0) \in \mathbb{R}^N$. The average consensus is said to be reached at
time $T$ if%
\begin{equation*}
x_{i}(k)=\bar{x}(0),i=1,2,\cdots ,N
\end{equation*}%
for any $k\geq T.$
\end{definition}

\begin{definition}
For a control sequence $\varepsilon (k)$, $0 \le k \le T-1$, in the form of (\ref{protocol}), the
corresponding graph filter is defined as
\begin{equation}
h(\lambda ,T):=\prod\limits_{k=0}^{T-1}(1-\varepsilon (k)\lambda ),
\label{profilter}
\end{equation}%
where $\lambda \in \mathbb{R}^N $ is the frequency variable.
\end{definition}

The graph filter defined by (\ref{profilter}) plays a key role in the analysis of MAS dynamics.
The following theorem shows that it characterizes
the necessary and sufficient condition of consensus.

\begin{theorem} \label{Th2}
For the MAS (\ref{agent}) on a connected graph $\mathcal{G}$ and under the control of sequence $\varepsilon (k)$ in (\ref{protocol}), let $%
0=\lambda _{1}<\lambda _{2}\leq \cdots \leq \lambda _{N}$ be the eigenvalues of the graph Laplacian matrix and $h({\lambda },T)$ be the graph filter defined by
(\ref{profilter}). Then \newline
\ \ \ (i) The MAS reaches average consensus at time $T<\infty $ if and only
if $h({\lambda _{i}},T)=0$ for $i=2,\cdots ,N.$\newline
\ \ \ (ii) The MAS reaches average consensus asymptotically if and only if
\begin{equation}
h({\lambda _{i}},\infty ):=\lim_{T\rightarrow \infty }h({\lambda _{i}}%
,T)=\lim_{T\rightarrow \infty }\left\{
\prod\limits_{k=0}^{T-1}(1-\varepsilon (k){\lambda _{i}})\right\} =0
\label{asymcond}
\end{equation}%
for $i=2,\cdots, N.$\newline
\ \ \ (iii) Assume that the control sequence satisfies $\lim_{k\rightarrow \infty}\varepsilon (k)=0$, and the MAS cannot reach consensus in finite-time. Then the consensus is reached asymptotically if and only if $%
\sum\limits_{k=0}^{\infty }{\varepsilon (k)}=\infty $.
\end{theorem}

\emph{Proof:} (i) It is easy to see that%
\begin{equation*}
x(T)=(I-{\varepsilon (T-1)}{\mathcal{L}})x(T-1)=\prod\limits_{k=0}^{T-1}(I-{\varepsilon
(k)}{\mathcal{L}})x(0)=\prod\limits_{k=0}^{T-1}(I-{\varepsilon (k)V\Lambda V}^{T})x(0).
\end{equation*}%
Recall that $\Lambda =diag\{0,\lambda _{2},\cdots,\lambda _{N}\},$ $V$ is unitary and $v_{1}=\frac{1}{N}{\vec{1}.}$ Therefore%
\begin{eqnarray}
x(T) &=&V\left[
\begin{array}{cccc}
1 &  &  &  \\
& \prod\limits_{k=0}^{T-1}(1-\varepsilon (k)\lambda _{2}) &  &  \\
&  & \ddots &  \\
&  &  & \prod\limits_{k=0}^{T-1}(1-\varepsilon (k)\lambda _{N})%
\end{array}%
\right] V^{T}x(0)  \notag \\
&=&Vdiag\{1,h(\lambda _{2},T),\cdots, h(\lambda _{N},T)\}V^{T}x(0)  \notag \\
&=&\bar{x}(0) {\vec{1}}%
+h(\lambda _{2},T)v_{2}v_{2}^{T}x(0)+\cdots +h(\lambda
_{N},T)v_{N}v_{N}^{T}x(0).  \label{xTeq}
\end{eqnarray}%
The `if' part is obvious from (\ref{xTeq}). Now suppose $h(\lambda
_{i},T)\neq 0$ for some $i.$ Setting $x(0)=v_{i},$ we know from the
orthoganality ${\vec{1}}^{T}{v}_{i}=0$ that the average of $v_{i}$ is zero.
However it follows from (\ref{xTeq}) that $x(T)=h(\lambda _{i},T)v_{i}$
is not zero. This contradiction proves the `only if' part.

(ii) It follows from (\ref{xTeq}) that
\begin{equation}
\lim_{T\rightarrow \infty }x(T)=\bar{x}(0) {\vec{1}}+\lim_{T\rightarrow \infty }\left(
h(\lambda _{2},T)v_{2}v_{2}^{T}+\cdots +h(\lambda
_{N},T)v_{N}v_{N}^{T}\right) x(0).  \label{proof2}
\end{equation}%
Then the `if' part follows from (\ref{proof2}) and the `only if' part can be proved by
contradiction similar to that in (i).

(iii) It follows from $\lim_{k\rightarrow \infty
}\varepsilon (k)=0$ that there exists $K>0$ such that $\varepsilon
(k)\lambda _{N}<1$ for any $k\geq K.$ Then we have $0<\varepsilon
(k)\lambda _{i}<1$ for $2\leq i\leq N $ and $k\geq K.$
Let $z(k)=\varepsilon (k)\lambda.$ Then using the fact that $1-z<e^{-z(k)}$ for any $z(k)>0$
and the assumption $\sum\limits_{k=0}^{\infty }{\varepsilon (k)}=\infty$, we have
\begin{equation}
\prod\limits_{k=K}^{\infty }(1-\varepsilon (k)\lambda )\leq e^{-\left(
\lambda \sum\limits_{k=K}^{\infty }\varepsilon (k)\right) }=0.
\end{equation}%
Therefore $h({\lambda _{i}},\infty )=0$ for $i=2,\cdots N,$ and, according to
(ii), the MAS reaches average consensus asymptotically.

We use contradiction to prove the `only if' part. Suppose that the MAS
system reaches average consensus asymptotically but not at any finite time
under a control protocol {satisfying }$\sum\limits_{k=0}^{\infty }{\varepsilon
(k)}<\infty .$ Note that ${\varepsilon (k)>0}$ and $\sum\limits_{k=0}^{%
\infty }{\varepsilon (k)}<\infty $ imply that $\lim_{K\rightarrow \infty
}\sum\limits_{k=K}^{\infty }{\varepsilon (k)=0.}$ Hence, there exists an integer
$K_{2}>0$ such that $\sum\limits_{k=K_{2}}^{\infty }\varepsilon (k)<\frac{1}{%
2\lambda _{N}}.$ Then for $0<\lambda \leq \lambda _{N},$ we have
\begin{equation*}
\prod\limits_{k=K_{2}}^{\infty }(1-\varepsilon (k)\lambda
)>1-\sum\limits_{k=K_{2}}^{\infty }{\varepsilon (k)\lambda >}\frac{1}{2}.
\end{equation*}%
Since the system cannot reach finite-time average consensus, we know from
(i) that there exists an $i,2\leq i\leq N$, such that $h({\lambda _{i}}%
,K_{2})\neq 0.$ Then $h({\lambda _{i}},\infty )=h({\lambda _{i}}%
,K_{2})\prod\limits_{k=K_{2}}^{\infty }(1-\varepsilon (k)\lambda _{i})\neq
0. $ This contradicts the result of (ii) that $h({\lambda _{i}}%
,\infty )=0$ for $i=2,\cdots ,N$ since it is supposed that the system
reaches consensus asymptotically. \hfill $\blacksquare $

\begin{remark}
Theorem 1 shows that the MAS reaches consensus if
and only if the graph filter $h(\lambda ,T)$ (or $h(\lambda ,\infty)$) is a `lowpass' filter which annihilates `high frequency' components at $\lambda_2, \cdots, \lambda_N$. In particular, we can set $\varepsilon(k)=\frac{1}{\lambda_{k+2}}$, $k=0,\cdots, N-2$, to achieve consensus at time $N-1$
since the corresponding graph filter
$h(\lambda ,N-1)=\prod\limits_{k=0}^{N-2}{(1-\frac{\lambda }{{\lambda _{k+2}}})}=0$ at $\lambda_i$, $i=2, \cdots, N$.
\end{remark}

\begin{remark}
The infinite product $h({\lambda _{i}}, \infty )=0$ is called `diverge to zero', which
isolates the case there is an exact zero term. The infinite product $h({\lambda _{i}}%
,\infty )=0$ not only provides novel insights into the asymptotic consensus, but also turns out
to be instrumental in the analysis of convergence rate as shown in the next section.
The Lyapunov based methods for asymptotic consensus usually provide sufficient conditions only
\cite{[23],[25],[26]}, which are somewhat conservative. To our knowledge, (\ref{asymcond})
is the simplest necessary and sufficient condition for asymptotic consensus that has established the
direct link between the control sequence and the graph topology.
\end{remark}

\begin{remark}
Time-varying control protocols with $\lim_{k\rightarrow \infty }\varepsilon
(k)=0$ are used in \cite{[9]}, where $\sum\limits_{k=0}^{\infty }{%
\varepsilon (k)}=\infty $ is an assumption, whereas we have shown here that it is actually a necessary condition.
\end{remark}

When the Laplacian matrix $\mathcal{L}$ has multiple eigenvalues,
the consensus time can be shorter than $N-1$. This is stated in the following corollary which follows directly from Theorem 1 (i).

\begin{corollary}
For a connected graph $\mathcal{G}$ with $N$ vertices, assume that its
Laplacian matrix has $K$ distinct nonzero eigenvalues $\lambda_{p_{k}}, k=0,\cdots ,K-1.$ Then $K$ is the minimum time for the MAS to reach
consensus. Moreover, by applying
\begin{equation}
{\varepsilon (k)=}\frac{1}{\lambda _{p_{k}}}, 0\leq k\leq K-1,
\label{contrfinite}
\end{equation}
the MAS (\ref{agent}) can reach average consensus at time $K$ to obtain $%
x(K)=\frac{1}{N}\sum\limits_{i=1}^{N}{{x_{i}}(0).}$
\end{corollary}

Based on Corollary 1, we summarize below some results of finite time consensus for the MAS on special graphs
(see Fig. 2 in Section \uppercase\expandafter{\romannumeral7} for demonstrative examples).

1) For a complete graph $\mathcal{G}$ with $N$ vertices, the
eigenvalues of its Laplacian matrix are
\begin{equation*}
{\lambda _{i}}=\left\{
\begin{array}{l}
0,\text{ }i=1 \\
M,\text{ }2\leq i\leq N
\end{array}%
\right. .
\end{equation*}%
Hence, the MAS can reach
consensus at the finite time $T=1$ by choosing ${\varepsilon (0)=}\frac{1}{N},$
that is,
\begin{equation*}
{u_{i}}(0)=\frac{1}{N}\sum\limits_{j\in {N_{i}}}{a_{ij}}({x_{j}}%
(0)-{x_{i}}(0)).
\end{equation*}

2) For a complete bipartite graph $\mathcal{G}$ with $M+N$ vertices, the
eigenvalues of its Laplacian matrix are
\begin{equation*}
{\lambda _{i}}=\left\{
\begin{array}{l}
0,\text{ }i=1 \\
M,\text{ }2\leq i\leq M-1 \\
N,\text{ }M\leq i\leq M+N-1 \\
M+N,\text{ }i=M+N%
\end{array}%
\right. .
\end{equation*}%
Hence, the MAS can reach consensus at the finite time $T=3$ by choosing the consensus protocol as follows.
\begin{eqnarray*}
{u_{i}}(0) &=&\frac{1}{M}\sum\limits_{j\in {N_{i}}}{{a_{ij}}({x_{j}}(0)-{%
x_{i}}(0))}, \\
{u_{i}}(1) &=&\frac{1}{N}\sum\limits_{j\in {N_{i}}}{{a_{ij}}({x_{j}}(1)-{%
x_{i}}(1))}, \\
{u_{i}}(2) &=&\frac{1}{M+N}\sum\limits_{j\in {N_{i}}}{{a_{ij}}({x_{j}}(2)-{%
x_{i}}(3))}.
\end{eqnarray*}

3) For a star graph $\mathcal{G}$ with $N$ vertices, the eigenvalues of its
Laplacian matrix are
\begin{equation*}
{\lambda _{i}}=\left\{
\begin{array}{l}
0,\text{ }i=1 \\
1,\text{ }2\leq i\leq N-1 \\
N,\text{ }i=N
\end{array}%
\right. .
\end{equation*}%
Hence, the MAS can reach
consensus at the finite time $T=2$ by choosing the consensus protocol as
\begin{eqnarray*}
{u_{i}}(0)&=&\sum\limits_{j\in {N_{i}}}{a_{ij}}({x_{j}}(0)-{x_{i}}(0)),\\
{u_{i}}(1)&=&\frac{1}{N}\sum\limits_{j\in {N_{i}}}{{a_{ij}}({x_{j}}(1)-{x_{i}}(1))}.
\end{eqnarray*}

4) For a cycle graph $\mathcal{G}$ with $N$ vertices, the eigenvalues of its
Laplacian matrix are ${\lambda _{1}}=0$, ${\lambda _{i}}=2-2\cos \frac{{%
2\pi (i-1)}}{N}$, $i=2, \cdots, floor(\frac{N+1}{2})$ with multiplicity 2,
and $\lambda_N=4$ with multiplicity 1 if $N$ is even.
Hence, the MAS can reach consensus at the finite time $T=ceil(\frac{N-1}{2})$ by
choosing ${\varepsilon (k)}=\frac{1}{\lambda _{k+2}},k=0, \cdots, floor(\frac{%
N+1}{2})-2$,
and ${\varepsilon(\frac{N}{2}-1)}=\frac{1}{4}$ if $N$ is even.

5) For a path graph $\mathcal{G}$ with $N$ vertices, the eigenvalues of its
Laplacian matrix are ${\lambda _{1}}=0$, ${\lambda _{i}}=2-2\cos \frac{{%
\pi (i-1)}}{N}$, $i=2, \cdots, N$.
Hence, the MAS can reach consensus at the finite time $T=N-1$ by
choosing ${\varepsilon (k)}=\frac{1}{\lambda _{k+2}},k=0, \cdots, N-2$.

\begin{remark}
The fact that finite time consensus can be achieved by choosing control
gains equal to the reciprocal of eigenvalues of the Laplacian matrix is not
new. It has been obtained by using different methods, for example, matrix
factorization method \cite{[30]}, minimal polynomial method \cite{[31]}, and
graph filter method \cite{[43]}.
However, by defining the graph filter $h(\lambda ,T)$,
Theorem 1 (i) and Corollary 1 have established the explicit connection between the
average consensus and filtering of graph signals.
\end{remark}

Corollary 1 has shown that finite time consensus can be achieved by the control strategy (\ref{contrfinite})
using the eigenvalue information of the graph Laplacian matrix.
This result not only gives a new interpretation of average consensus, but also provides a systematic method to characterize the convergence rate explicitly. In particular, we can show that the exact convergence rate is equal to $\rho =\max_{\{\lambda _{i}\}}{\left\vert {h({\lambda _{i}}, \cdot)}\right\vert }$ for periodic control sequences. Following this, we can further convert the fast consensus design problem to a polynomial interpolation problem, which can be readily solved by various methods. The polynomial interpolation problem will be discussed in section V in detail.

\section{Exact Convergence Rate of Periodic Control Sequences}

In this section, we analyze the convergence rate of asymptotic consensus. We focus on the
$M$-periodic control sequence ${\varepsilon }(k+M)={\varepsilon }(k)$, $\forall k \in \mathbb{Z}$.
Before presenting the main result, we introduce some definitions.

For an $M$-periodic control sequence ${\varepsilon }(k)$, define the corresponding graph filter as
\begin{equation}\label{pericontrol}
h(\lambda ,M)=\prod\limits_{k=0}^{M-1}{(1-{\varepsilon (k)}{\lambda }).}
\end{equation}%
The error of consensus at time $k$ is defined as $e(k):=x(k)-\bar{x}(0){\vec{1}}$.
As an extension of per-step convergence rate in \cite{[16]}, the per-period convergence rate for an
$M$-periodic control sequence ${\varepsilon }(k)$ is defined as
\begin{equation}
\rho_M := \mathop {\sup }\limits_{\left\Vert{x(0)}\right\Vert _{2} = 1} \frac{\left\Vert{e(M)}\right\Vert _{2}}{\left\Vert{e(0)}\right\Vert _{2}}.  \label{convrate}
\end{equation}

\begin{theorem}
For the MAS (\ref{agent}) on a connected graph $\mathcal{G}$ and under the control of an $M$-periodic control sequence
${\varepsilon }(k)$, let $%
0=\lambda _{1}<\lambda _{2}\leq \cdots \leq \lambda _{N}$ be the eigenvalues
of the graph Laplacian matrix and $h(\lambda, M)$ be as given by (\ref{pericontrol}).
Then the MAS reaches consensus asymptotically if and only if
the exact convergence rate $\rho^{*} <1$, where
\begin{equation}\label{rhostar}
\rho^{*} := \max_{\{\lambda _{i}\}}\left\vert {h({\lambda _{i}}, M)}\right\vert.
\end{equation}
Moreover, the per-period convergence rate $\rho_M$ equals the exact convergence rate $\rho^{*}$, that is,
\begin{equation}\label{rhoM}
\rho_M = \mathop {\sup }\limits_{\left\Vert{x(0)}\right\Vert _{2} = 1} \frac{\left\Vert{e(M)}\right\Vert _{2}}{\left\Vert{e(0)}\right\Vert _{2}} = \max_{\{\lambda _{i}\}}\left\vert {h({\lambda _{i}}, M)}\right\vert =\rho^*.
\end{equation}
\end{theorem}

\begin{IEEEproof}
Since ${\varepsilon (k)}$ is $M$-periodic, we have
\begin{eqnarray}
h(\lambda ,jM) &=&\prod\limits_{k=0}^{jM-1}(1-\varepsilon (k)\lambda )=\left(
\prod\limits_{k=0}^{M-1}(1-\varepsilon (k)\lambda )\right) ^{j}=\left(
h(\lambda ,M)\right) ^{j} 
\end{eqnarray}%
It follows from $\left\vert {h({\lambda_i}, jM)}\right\vert \leq (\rho^*)^j$
and $\rho^{*} < 1$ that
$\lim_{j\rightarrow \infty }h({\lambda _{i}},jM+l)=0$ for $l=0,\cdots,
M-1,i=2,\cdots ,N.$ Then the average consensus is guaranteed by the `if'
part of  Theorem 1(ii). On the other hand, if the MAS reaches average consensus, we
know from Theorem 1(ii) that $\lim_{j\rightarrow \infty }h({\lambda _{i}}%
, jM)=\lim_{j\rightarrow \infty }\left( h(\lambda_i, M)\right) ^{j}=0$ for $%
i=2,\cdots ,N.$ This holds only if $\rho^{*}=\max_{\{\lambda _{i}\}}\left\vert {h({\lambda _{i}}, M)}\right\vert < 1.$

By using (\ref{xTeq}) and the unitariness of $V$, we have
\begin{eqnarray}\label{rholeq}
\rho_M &=& \mathop {\sup }\limits_{\left\Vert{x(0)}\right\Vert _{2} = 1} \frac{\left\Vert{x(M)-\bar{x}(0){\vec{1}}}\right\Vert _{2}}{\left\Vert{x(0)-\bar{x}(0){\vec{1}}}\right\Vert _{2}} \nonumber \\
&=& \mathop {\sup }\limits_{{{\left\Vert {x(0)} \right\Vert}_2} = 1} \frac{{\left\Vert {\sum\limits_{i = 2}^N {h({\lambda _i},M){v_i}v_i^Tx(0)} } \right\Vert_2}}{{\left\Vert {\sum\limits_{i = 2}^N {{v_i}v_i^Tx(0)} } \right\Vert_2}} \nonumber \\
&\leq& \mathop {\sup }\limits_{{{\left\Vert {x(0)} \right\Vert}_2} = 1} \frac{{\mathop {\max }\limits_{\{\lambda _i\}} \left| {h({\lambda _i},M)} \right|{{\left\| {\sum\limits_{i = 2}^N {{v_i}v_i^Tx(0)} } \right\|}_2}}}{{{{\left\| {\sum\limits_{i = 2}^N {{v_i}v_i^Tx(0)} } \right\|}_2}}} \nonumber \\
&=& \mathop {\max }\limits_{\{\lambda_i\}} \left\vert {h({\lambda _i},M)} \right\vert = \rho^{*}.
\end{eqnarray}

Assume that $\lambda_j$ is the eigenvalue such that $\left\vert {h({\lambda _j},M)} \right\vert = \mathop {\max }\limits_{\{\lambda_i\}} \left\vert {h({\lambda _i},M)} \right\vert$.
Setting $x(0)=v_j$, we have $\left\Vert v_j \right\Vert_2 = 1$, $v_j {\vec{1}} =0$ and
$x(M)=h(\lambda_j, M)v_j$. Hence,
\begin{equation}
\frac{\left\Vert{e(M)}\right\Vert _{2}}{\left\Vert{e(0)}\right\Vert _{2}} = \frac{\left\Vert{h(\lambda_j, M)v_j}\right\Vert _{2}}{\left\Vert{v_j}\right\Vert _{2}} = \left\vert {h({\lambda _j},M)} \right\vert.
\end{equation}
From the definition of (\ref{convrate}), we have
\begin{equation}\label{rhogeq}
\rho_M \geq \left\vert {h({\lambda _j},M)} \right\vert.
\end{equation}
Therefore, (\ref{rhoM}) is proved by combining (\ref{rholeq}) and (\ref{rhogeq}).
\end{IEEEproof}

\begin{remark}
The $\rho_M$ in (\ref{rhoM}) represents the exact convergence rate in the sense that
$\rho_M \leq \gamma$ for any $\gamma$ satisfying
$\left\Vert{e(jM)}\right\Vert _{2} \leq \gamma^{j}\left\Vert e(0)\right\Vert_2$.
Whereas many results based on Lyapunov function \cite{[9],[12],[09],[10]} can only give the
upper bound of the convergence rate, which is usually much larger than $\rho_M$.
\end{remark}

\begin{remark}
Theorem 2 turns the problem of finding a faster consensus algorithm to
the problem of designing a polynomial $h(\lambda, M)$ such that $\rho_M$ (or $\rho ^{*}$)
in (\ref{rhoM}) is small. Various polynomial
interpolation techniques can be employed to solve this problem. We will explore this issue in detail
in Sections V and VI.
\end{remark}

\section{Consensus on Uncertain Networks by Lagrangian Polynomial Interpolation}

In practical applications, it is usually difficult, if not impossible, to obtain the exact eigenvalues of the Laplacian matrix, especially when the network is large and complex.
However, there are efficient methods to estimate $\lambda_2$ and $\lambda_N$ \cite{[50]}.
In this section, we consider the MAS on uncertain networks with $[\lambda_2, \lambda_N] \subseteq [\alpha, \beta]$,
where $\alpha$ is the lower bound of the algebraic connectivity and $\beta$ is the upper bound of the Laplacian
spectral radius.

Define the worst-case convergence rate  as
\begin{equation}\label{gamma}
{\gamma}_{M} := \mathop {\sup }\limits_{\{\mathcal{G}\}_{[\alpha, \beta]}} \rho_M,
\end{equation}
where $\rho_M$ is the per-period convergence rate defined in (\ref{convrate}), and $\{\mathcal{G}\}_{[\alpha, \beta]}$ is defined as the set of all connected graphs with $[\lambda_2, \lambda_N] \subseteq [\alpha, \beta]$. Then the following lemma is immediate.
\begin{lemma}
The worst-case convergence rate satisfies
\begin{equation}\label{gammaM}
{\gamma}_{M} = \mathop {\sup }\limits_{\{\mathcal{G}\}_{[\alpha, \beta]}} \rho^* = \max_{\lambda \in {[\alpha, \beta]}}\left\vert {h({\lambda}, M)}\right\vert,
\end{equation}
where $\rho^*$ is the exact convergence rate defined in (\ref{rhostar}).
\end{lemma}

If we assume that the eigenvalues of the graph Laplcaian are distributed uniformly on $[\alpha, \beta]$,
then it is natural to choose an $M$-periodic control sequence by distributing $\frac{1}{\varepsilon (k)}$,
$k=0,1,\cdots,M-1$, uniformly on $[\alpha, \beta]$,
and use Lagrangian polynomial interpolation to obtain the control sequence.

Choose $M$ points in $[\alpha ,\beta]$ as
\begin{equation}\label{r-k}
{r}_{k}=\alpha +\frac{\beta -\alpha }{M+1}(k+1), k=0,\cdots ,M-1,
\end{equation}%
and set the $M$-periodic control sequence as
\begin{equation}
\varepsilon (k+j{M})=\frac{1}{r_k}=\frac{1}{\alpha +\frac{\beta -\alpha }{M+1}(k+1)},%
k=0, \cdots, {M-1}, j =0, 1, \cdots.  \label{Equ_gain}
\end{equation}
Theorem 3 below shows that the $M$-periodic control sequence (\ref{Equ_gain}) does provide consensus.
It also presents an explicit expression of the worst-case convergence rate $\rho_M$,
which asserts the better performance of (\ref{Equ_gain}) as compared to the existing result \cite{[16]}.

\begin{theorem}\label{Th3}
For any MAS (\ref{agent}) on a connected graph ${\mathcal{G}} \in \{\mathcal{G}\}_{[\alpha, \beta]}$,
let the control sequence
be given by (\ref{Equ_gain}) with the corresponding graph filter
\begin{equation}
h(\lambda, {M})=\prod\limits_{k=0}^{M-1}{\left( 1-\frac{\lambda }{{r}%
_{k}}\right) }  \label{eq23}
\end{equation}
and $r_k$ given by (\ref{r-k}).
Then the MAS reaches average consensus and the per-period convergence rate satisfies
$\rho_M \leq \gamma_M < 1$, where
\begin{equation}
\gamma_M =\frac{M!}{\prod\limits_{k=1}^{M}(k+\frac{M+1}{\beta -\alpha }\alpha )}. \label{Equ_rho}
\end{equation}
\end{theorem}

\begin{IEEEproof}
For ${\lambda }\in \lbrack \alpha +\frac{\beta -\alpha }{M+1}i,\alpha +\frac{%
\beta -\alpha }{M+1}(i+1)],i=0,\cdots ,M,$ we have from (\ref{eq23})%
\begin{eqnarray*}
\left\vert h(\lambda, {M})\right\vert  &=&\prod\limits_{k=0}^{i-1}{\left(
\frac{\lambda }{{r}_{k}}-1\right) \cdot }\prod\limits_{k=i}^{M-1}{%
\left( 1-\frac{\lambda }{{r}_{k}}\right) } \\
&\leq &\prod\limits_{k=0}^{i-1}\frac{\alpha +\frac{\beta -\alpha }{M+1}%
(i+1)-\left( \alpha +\frac{\beta -\alpha }{M+1}(k+1)\right) }{\alpha +\frac{%
\beta -\alpha }{M+1}(k+1)}{\cdot }\prod\limits_{k=i}^{M-1}\frac{\alpha +\frac{%
\beta -\alpha }{M+1}(k+1)-\left( \alpha +\frac{\beta -\alpha }{M+1}i\right) }{%
\alpha +\frac{\beta -\alpha }{M+1}(k+1)} \\
&=&\prod\limits_{k=0}^{i-1}\frac{\frac{\beta -\alpha }{M+1}(i-k)}{\alpha +%
\frac{\beta -\alpha }{M+1}(k+1)}{\prod\limits_{k=i}^{M-1}\frac{\frac{\beta
-\alpha }{M+1}(k+1-i)}{\alpha +\frac{\beta -\alpha }{M+1}(k+1)k}} \\
&=&\prod\limits_{k=0}^{i-1}\frac{(i-k)}{(k+1+\frac{M+1}{\beta -\alpha }\alpha )%
}{\prod\limits_{k=i}^{M-1}}\frac{(k+1-i)}{(k+1+\frac{M+1}{\beta -\alpha }\alpha )%
} \\
&=&\frac{i!(M-i)!}{\prod\limits_{k=1}^{M}(k+\frac{M+1}{\beta -\alpha }\alpha
)}
\leq \frac{M!}{\prod\limits_{k=1}^{M}(k+\frac{M+1}{\beta -\alpha }\alpha )} = \gamma_M 
\end{eqnarray*}
and the equality holds if and only if $\lambda=\alpha$ or $\lambda=\beta$.
It then follows from (\ref{rhoM}) that
\begin{equation*}
\rho_M = \max_{\{\lambda _{i}\}}\left\vert {h({\lambda _{i}}, M)}\right\vert \leq
\mathop {\max }\limits_{\lambda  \in [\alpha ,\beta ]} \left\vert h(\lambda, {M})\right\vert
\leq \frac{M!}{\prod\limits_{k=1}^{M}(k+\frac{M+1}{\beta -\alpha }\alpha )}%
=\gamma_M
\end{equation*}
and $\gamma_M <1$ since $\frac{M+1}{\beta -\alpha }\alpha > 0$.
By Theorem 2, we know that the MAS reaches average consensus asymptotically.
\end{IEEEproof}

\begin{remark}\label{Remark7}
In \cite{[16]}, it is proved that the fastest convergence rate by constant control
sequence is $\rho=\frac{{\beta - \alpha}}{{\beta + \alpha}}$, with
$\varepsilon(k) \equiv {\varepsilon_c} = \frac{2}{{{\alpha} + {\beta}}}$. This is a special case of $M=1$ in (\ref{Equ_gain}).
Because
\begin{equation}
\gamma_M  = \frac{{M!}}{{\prod\limits_{k = 1}^M {(k + \frac{{M + 1}}{{\beta  - \alpha }}\alpha )} }} = \prod\limits_{k = 1}^M {\frac{{k(\beta  - \alpha )}}{{k(\beta  - \alpha ) + (M + 1)\alpha }}}  = \prod\limits_{k = 1}^M {\frac{{\beta  - \alpha }}{{\beta  + \frac{{M + 1 - k}}{k}\alpha }}}
\end{equation}
and $\prod\limits_{k = 1}^M {(\beta  + \frac{{M + 1 - k}}{k}\alpha ) > (} \beta  + \alpha {)^M}$, we have
\begin{equation*}
\gamma_M {\leq }\left( \frac{\beta -\alpha }{\beta +\alpha} \right) ^{M}={\left(1-%
\frac{2\alpha }{\beta+\alpha}\right)^{M}}
\end{equation*}%
and the equality holds if and only if $M=1$. Therefore, for $M>1$, the convergence performance of $M$-periodic control sequence (\ref{Equ_gain}) is better than that of the fastest constant control gain in \cite{[16]}.
\end{remark}

When the algebraic connectivity $\alpha$ and the Laplacian spectral radius $\beta$ are uncertain, we have the following result.

\begin{corollary}
For the MAS (\ref{agent}) on an uncertain connected graph ${\mathcal{G}}$, set the $M$-periodic control sequence
as
\begin{equation}
{\varepsilon ({k+j{M}})}=\frac{M+1}{{\bar\beta}k}, {k=0, 2, \cdots,{M-1}},
{j=0, 1, \cdots},  \label{eq24}
\end{equation}%
where ${\bar\beta} > \beta$ is a constant.
Then the MAS reaches average consensus asymptotically.
Furthermore, if $M$ is large enough such that all the nonzero eigenvalues of the graph Laplacian are located
in the interval $\lbrack \frac{{\bar\beta}}{M+1}, \frac{M{\bar\beta}}{M+1}\rbrack$, then the convergence rate
satisfies $\rho_M \leq \gamma_M = \frac{1}{M}$.
\end{corollary}
\begin{IEEEproof}
For the $M$-periodic control gain in (\ref{eq24}), the corresponding graph filter can be obtained as
$h(\lambda, {M})=\prod\limits_{k=0}^{M-1}{\left( 1-\frac{(M+1)\lambda }{{\bar\beta} (k+1)}\right) }$.
For ${\lambda }\in \lbrack \frac{{\bar\beta}}{M+1}i, \frac{{\bar\beta}}{M+1}(i+1)\rbrack, {\lambda } \ne 0, {\lambda } \ne {\bar\beta},
i=0,\cdots ,M,$ we have%
\begin{eqnarray*}
\left\vert h(\lambda ,{M})\right\vert  &=&\prod\limits_{k=0}^{i-1}{\left(
\frac{(M+1)\lambda }{{\bar\beta}(k+1)}-1\right) \cdot }\prod\limits_{k=i}^{M-1}{%
\left( 1-\frac{(M+1)\lambda }{{\bar\beta}(k+1)}\right) } \\
&\leq &\prod\limits_{k=0}^{i-1}\frac{{\bar\beta}(i+1)-{\bar\beta}(k+1)}{{\bar\beta} (k+1)}{\cdot }\prod\limits_{k=i}^{M-1}\frac{{\bar\beta}(k+1)-{\bar\beta}i }{{\bar\beta}(k+1)} \\
&=&\prod\limits_{k=0}^{i-1}\frac{i-k}{k+1}{\prod\limits_{k=i}^{M-1}%
\frac{k+1-i}{k+1}} \\
&=&\frac{i!(M-i)!}{M!}.
\end{eqnarray*}
Thus
\begin{equation*}
\rho_M \leq \mathop {\max }\limits_{\lambda  \in (0, \bar\beta)} \left\vert h(\lambda, {M})\right\vert <
\left\vert h(0, {M})\right\vert = \left\vert h({\bar\beta}, {M})\right\vert=1.
\end{equation*}
From Theorem 2, we know that the MAS reaches average consensus asymptotically.

For ${\lambda }\in \lbrack \frac{{\bar\beta}}{M+1}, \frac{M{\bar\beta}}{M+1}\rbrack$, it is easy to verify that
\begin{equation*}
\gamma_M = {\max _{i \in \{ 1, \cdots ,M - 1\} }}\frac{{i!(M - i)!}}{{M!}}=\frac{1}{M}.
\end{equation*}
It follows from Theorem 2 that $\rho_M = \max_{\{\lambda _{i}\}}\left\vert {h({\lambda _{i}}, M)}\right\vert \leq \gamma_M = \frac{1}{M}$.
\end{IEEEproof}

\section{Explicit Solutions of the Optimal Consensus on Uncertain Networks}

In the last section, we have presented an intuitive periodic control sequence (\ref{Equ_gain}) to achieve consensus and have presented the explicit expression of the worst-case convergence rate for uncertain networks in the set $\{\mathcal{G}\}_{[\alpha, \beta]}$. In this section, we study the optimal design problem to find the
$M$-periodic control sequence $\varepsilon (k)$ that yields the
optimal solution of the worst-case convergence rate $\gamma_M$ for any graph
$\mathcal{G} \in \{\mathcal{G}\}_{[\alpha, \beta]}$. Precisely, we
will solve the following polynomial interpolation problem
\begin{equation}
\min_{\varepsilon (k): \varepsilon (k+M)=\varepsilon (k)}{\gamma_M} =
\min_{\varepsilon (k), k=0,\cdots,M-1}\max_{\lambda \in \left[ \alpha ,\beta \right]
}\left\vert {\prod\limits_{k=1}^{M}(1-\varepsilon (k)\lambda)}
\right\vert  \label{polyopt}
\end{equation}
to find the optimal value $\gamma^*_M$ and the optimal $M$-periodic control sequence $\{\varepsilon^*(k)\}$,
where $\gamma_M$ is defined in (\ref{gamma}).

In \cite{[51]}, the authors have used Chebyshev polynomial as the orthogonal bases to
construct the least-squares approximation $h(\lambda)=\Sigma _{i=0}^{M}h_{i}\lambda ^{i}$ for a desired filter. However, different from \cite{[51]}, the problem (\ref{polyopt}) is a uniform interpolation problem that cannot be solved by orthogonal basis approximation. A possible fix to this problem is to solve (\ref{polyopt}) numerically by linear programming.
Besides computational complexity, this numerical approach suffers from at least two drawbacks. One is that we cannot compute
$\gamma^*_M$ directly from $\alpha, \beta, M$, and the essential relation between
$\gamma^*_M$ and $\alpha, \beta, M$ is hidden. The other is that the optimal polynomial thus computed might have complex roots, leading to complex valued $\varepsilon (k)$. This cannot be implemented by the control protocol (\ref{protocol}), which requires $\varepsilon (k) > 0$. In fact, all the polynomial approximation based methods in the literature \cite{[21],[22],[43],[44],[45]} cannot guarantee the designed polynomial to have only positive real roots.

To overcome the above difficulties, we will use Chebyshev polynomial interpolation to derive the explicit and analytical solution to the problem (\ref{polyopt}). We first construct the polynomial $h(\lambda, M)$, and then prove that it is the unique and optimal solution of (\ref{polyopt}).

For ${\chi }\in \lbrack -1,1]$, the Chebyshev polynomials $\mathcal{T}_{M}({%
\chi })$, $M=0,1,2,\cdots $, are defined as \cite{[46],[52]}
\begin{equation}
{\mathcal{T}_{M}}({\chi }):=\left\{ {%
\begin{array}{cc}
{1,} & {M=0} \\
{\chi ,} & {M=1} \\
{2\chi {\mathcal{T}_{M-1}}(\chi )-{\mathcal{T}_{M-2}}(\chi ),} & {M\geq 2}%
\end{array}%
}\right..  \label{eq15}
\end{equation}%
The following properties hold for Chebyshev polynomials $\{\mathcal{T}%
_{M}({\chi })\}$ \newline
(i) ${\mathcal{T}_{M}}({\chi })=\cos (M\arccos {\chi })$ for ${\chi }\in
\lbrack -1,1],$ \newline
(ii) ${\mathcal{T}_{M}}(\cos {\frac{{2i-1}}{{2M}}\pi })=0$ for $i=1,2,\cdots
,M ,$ \newline
(iii) ${\max_{x\in \lbrack -1,1]}}\left\vert {{\mathcal{T}_{M}}(\chi )}%
\right\vert =1$, and ${\mathcal{T}_{M}}({\chi })= \pm 1$ alternately at
${\chi }_{i}=\cos {\frac{i}{M}\pi }$, $i=0,1,\cdots ,M$.

Setting $\lambda =\frac{\beta -\alpha }{2}{\chi }+\frac{%
\beta +\alpha }{2}$ in $\mathcal{T}_{M}({\chi }),$ we get a new polynomial $%
g_{M}(\lambda )$ on $\lambda \in \lbrack \alpha ,\beta ]$
\begin{equation}
g_{M}(\lambda ):=\mathcal{T}_{M}({\chi })=\cos \left[ M\arccos \left( {\frac{2}{\beta
-\alpha }}\lambda -{\frac{\beta +\alpha }{\beta -\alpha }}\right) \right] .
\label{eq16}
\end{equation}
Based on the relation between $\chi $ and $\lambda $, Tabel \ref{ChebyPoly} can
be derived from (\ref{eq15}) and (\ref{eq16}).
\begin{table}
\caption{\label{ChebyPoly}Chebyshev Polynomials}\centering
\begin{tabular}{c|c|c}
\hline
$M$ & $\mathcal{{T}_{M}(\chi )}$, $\chi \in \lbrack -1,1]$ & $g_{M}(\lambda )$%
, $\lambda \in \lbrack \alpha ,\beta ]$ \\ \hline\hline
$M=1$ & $\chi $ & ${\frac{2}{\beta -\alpha }}\lambda -{\frac{\beta +\alpha }{%
\beta -\alpha }}$ \\ \hline
$M=2$ & $2\chi ^{2}-1$ & $2\left( {\frac{2}{\beta -\alpha }}\lambda -{\frac{%
\beta +\alpha }{\beta -\alpha }}\right) ^{2}-1$ \\ \hline
$M=3$ & $4\chi ^{3}-3\chi $ & $4\left( {\frac{2}{\beta -\alpha }}\lambda -{%
\frac{\beta +\alpha }{\beta -\alpha }}\right) ^{3}-3\left( {\frac{2}{\beta
-\alpha }}\lambda -{\frac{\beta +\alpha }{2{\beta }-\alpha }}\right) $ \\
\hline
$M=4$ & $8\chi ^{4}-8\chi ^{2}+1$ & $8\left( {\frac{2}{\beta -\alpha }}%
\lambda -{\frac{\beta +\alpha }{\beta -\alpha }}\right) ^{4}-8\left( {\frac{2%
}{\beta -\alpha }}\lambda -{\frac{\beta +\alpha }{\beta -\alpha }}\right)
^{2}+1$ \\ \hline
$M=5$ & $16\chi ^{5}-20\chi ^{3}+5\chi $ & $16\left( {\frac{2}{\beta -\alpha
}}\lambda -{\frac{\beta +\alpha }{\beta -\alpha }}\right) ^{5}-20\left( {%
\frac{2}{\beta -\alpha }}\lambda -{\frac{\beta +\alpha }{\beta -\alpha }}%
\right) ^{3}+5\left( {\frac{2}{\beta -\alpha }}\lambda -{\frac{\beta +\alpha
}{\beta -\alpha }}\right) $ \\ \hline
$M=6$ & $32\chi ^{6}-48\chi ^{4}+18\chi ^{2}-1$ & $32\left( {\frac{2}{\beta
-\alpha }}\lambda -{\frac{\beta +\alpha }{\beta -\alpha }}\right)
^{6}-48\left( {\frac{2}{\beta -\alpha }}\lambda -{\frac{\beta +\alpha }{%
\beta -\alpha }}\right) ^{4}+18\left( {\frac{2}{\beta -\alpha }}\lambda -{%
\frac{\beta +\alpha }{\beta -\alpha }}\right) ^{2}-1$ \\ \hline
\end{tabular}%
\end{table}

\begin{lemma}
\bigskip The polynomial $g_{M}(\lambda )$ has the following
properties:\newline
(i) $g{_{M}}({\frac{\beta -\alpha }{2}}{\cos {\frac{{2i-1}}{{2M}}\pi }}+%
\frac{\beta +\alpha }{2})=0$ for $i=1,2,\cdots ,M$.\newline
(ii) $\max_{\lambda \in \lbrack \alpha ,\beta ]}|g_{M}(\lambda )|=1$ and $%
g_{M}(\lambda _{i})=\pm 1$ alternatively at $\lambda _{i}=\frac{\beta
-\alpha }{2}{\chi }+\frac{\beta +\alpha }{2},$ $i=0,1,\cdots M.$\newline
(iii) $g_{M}(0)$ can be written as
\begin{equation}
g_{M}(0)=\frac{1}{2}(-1)^{M}\left( \frac{\sqrt{\beta /\alpha }-1}{\sqrt{%
\beta /\alpha }+1}\right) ^{M}+\frac{1}{2}(-1)^{M}\left( \frac{\sqrt{\beta
/\alpha }+1}{\sqrt{\beta /\alpha }-1}\right) ^{M}, \label{gm0}
\end{equation}
and $\underset{M\rightarrow \infty }{\lim }\left\vert g_{M}(0)\right\vert=\infty$ monotonically.
\end{lemma}

\begin{IEEEproof}
(i) and (ii) follow immediately from the relation $\lambda =\frac{\beta -\alpha }{2}%
{\chi }+\frac{\beta +\alpha }{2}$. We now prove (iii). Note that $g_{M}(\lambda )$ can be written as
\begin{equation}
g{_{M}}(\lambda )=\left\{ {%
\begin{array}{cc}
{1,} & {M=0} \\
{\frac{2}{{\beta -\alpha }}\lambda -\frac{{\beta +\alpha }}{{\beta -\alpha }}%
,} & {M=1} \\
{2\left( {\frac{2}{{\beta -\alpha }}\lambda -\frac{{\beta +\alpha }}{{\beta
-\alpha }}}\right) g{_{M-1}}(\lambda )-g{_{M-2}}(\lambda ),} & {M\geq 2}%
\end{array}%
}\right..   \label{eq19}
\end{equation}

Hence we have
\begin{equation}
g{_{M}(0)}=\left\{ {%
\begin{array}{cc}
{1,} & {M=0} \\
{-\frac{{\beta +\alpha }}{{\beta -\alpha }},} & {M=1} \\
{{-\frac{2({\beta +\alpha )}}{{\beta -\alpha }}}g{_{M-1}(0)}-g{_{M-2}(0)},}
& {M\geq 2}%
\end{array}%
}\right..   \label{gm0iteration}
\end{equation}

By direct verification, ${g}_{M}(0)$ in (\ref{gm0}) is the explicit formula of
iteration (\ref{gm0iteration}). It follows directly from (\ref{gm0}) that
\begin{equation}
\left\vert g_{M}(0)\right\vert =\frac{1}{2}\left( 1-\frac{2}{\sqrt{\beta
/\alpha }+1}\right) ^{M}+\frac{1}{2}\left( 1+\frac{2}{\sqrt{\beta /\alpha }-1}\right) ^{M}.
\end{equation}%
Then
\begin{equation}
\nonumber
\begin{array}{l}
\left| {{g_{M + 1}}(0)} \right| - \left| {{g_M}(0)} \right|\\
 = \frac{1}{2}{\left( {1 - \frac{2}{{\sqrt {\beta /\alpha }  + 1}}} \right)^{M + 1}} + \frac{1}{2}{\left( {1 + \frac{2}{{\sqrt {\beta /\alpha }  - 1}}} \right)^{M + 1}} - \frac{1}{2}{\left( {1 - \frac{2}{{\sqrt {\beta /\alpha }  + 1}}} \right)^M} - \frac{1}{2}{\left( {1 + \frac{2}{{\sqrt {\beta /\alpha }  - 1}}} \right)^M}\\
 = \frac{1}{2}{\left( {1 + \frac{2}{{\sqrt {\beta /\alpha }  - 1}}} \right)^M}\frac{2}{{\sqrt {\beta /\alpha } - 1}} - \frac{1}{2}{\left( {1 - \frac{2}{{\sqrt {\beta /\alpha } + 1}}} \right)^M}\frac{2}{{\sqrt {\beta /\alpha }  + 1}}\\
 > \frac{1}{{\sqrt {\beta /\alpha } - 1}} - \frac{1}{{\sqrt {\beta /\alpha } + 1}}\\
 = \frac{2}{{\beta /\alpha } - 1} > 0.
\end{array}
\end{equation}
Therefore the sequence $\left\vert g_{M}(0)\right\vert $ approaches infinity
monotonically as $M$ tends to infinity.
\end{IEEEproof}

\begin{theorem}\label{Th4}
Define $h(\lambda ,M)=\frac{1}{g{_{M}(0)}}g{_{M}}(\lambda )$. Then $%
h(\lambda ,M)$ is the unique $M$-th order polynomial, with $M$ roots
\begin{equation}\label{ri}
r_i = {\frac{\beta -\alpha }{2}}{\cos {\frac{{2i-1}}{{2M}}\pi }}+\frac{\beta +\alpha }{2}, i=1,2,\cdots,M,
\end{equation}
that solves (\ref{polyopt}) and yields the optimal worst-case convergence rate $\gamma^*_M=\frac{1}{\left\vert g_{M}(0)\right\vert}$. The $h(\lambda, M)$ can be rewritten as
\begin{equation}\label{chebyh}
h(\lambda, M)=\overset{M-1}{\underset{k=0}\Pi}\left(1-\frac{1}{r_{k+1}}\lambda\right)
\end{equation}%
with $h(0,M)=1$.
\end{theorem}

\begin{IEEEproof}
From Lemma 2, $h(\lambda, M)$ is an $M$-th order polynomial with roots as given in (\ref{ri}). Hence, it can be written as (\ref{chebyh}). Since $\max_{\lambda \in \lbrack \alpha ,\beta \rbrack}|g_{M}(\lambda )|=1$ and $%
g_{M}(\lambda _{i})=\pm 1$ alternatively at $\lambda _{i}=\frac{\beta
-\alpha }{2}{\chi }+\frac{\beta +\alpha }{2},$ $i=0,1,\cdots M$,
the polynomial $h(\lambda, M)$ satisfies
$\max_{\lambda \in \lbrack \alpha ,\beta \rbrack}|h(\lambda, M)|=\frac{1}{{\left\vert g_{M}(0)\right\vert}}$
and $h(\lambda, M)=\pm \frac{1}{{\left\vert g_{M}(0)\right\vert}}$ alternatively at $M+1$ points in $[\alpha, \beta]$.
By Chebyshev alternation theorem (p. 30, \cite{[52]}), we known that this $h(\lambda, M)$ is the unique $M$-th order polynomial
that solves the optimal interpolation problem (\ref{polyopt}). Finally, the optimality of $\gamma^*_M=\frac{1}{\left\vert g_{M}(0)\right\vert}$ is obvious.
\end{IEEEproof}

Combining Theorems 2 and \ref{Th4}, we immediately obtain the following result.

\begin{theorem}
For any MAS (\ref{agent}) on a connected graph $\mathcal{G} \in \{\mathcal{G}\}_{[\alpha, \beta]}$,
set the $M$-periodic control sequence as
\begin{equation}
{\varepsilon^* ({k+jM})}=\frac{1}{r_{k+1}},k=0,1,\cdots ,M-1, j=0,1,\cdots,
\label{eq26}
\end{equation}%
where $r_k$ is defined in (\ref{ri}).
Then the MAS reaches average consensus asymptotically and the exact convergence rate
$\rho_M$ satisfies
\begin{equation}
\rho_M \leq \gamma^*_M = \frac{1}{\left\vert g_{M}(0)\right\vert}.
\end{equation}%
Moreover, for any other $M$-periodic
control sequence $\varepsilon (k)$, there always exists a connected graph
$\tilde {\mathcal{G}} \in \{\mathcal{G}\}_{[\alpha, \beta]}$
such that its convergence rate under the $\varepsilon (k)$ is $\rho_{\varepsilon (k)} > \gamma^*_M$.
\end{theorem}

\begin{remark}
Although the proof is simple, the implication of Theorem 5 is important. It means that $\gamma^*$ is the fastest convergence rate for the worst-case scenario if $\alpha$ and $\beta$ are the only information we know about the network.
$\gamma^*_M=\frac{1}{{\left\vert g_{M}(0)\right\vert}}$ also provides the direct relation between the bound of convergence rate and the graph topology.
It is obvious that $\gamma^*_M  \approx 2\left( \frac{\sqrt {\beta /\alpha }-1}{\sqrt {\beta /\alpha }+1}\right)^M$ for large $M$.
\end{remark}

Next, we present the performance limitation of the optimal worst-case convergence rate under the general time-varying control sequences.

For a given graph $\mathcal{G}$, the asymptotic convergence rate is defined as
\begin{equation}\label{rhostep}
\rho_{asym}:= \mathop {\sup }\limits_{\left\Vert{x(0)}\right\Vert _{2} = 1} \underset{M\rightarrow \infty }{\lim } \left( \frac{\left\Vert{e(M)}\right\Vert _{2}}{\left\Vert{e(0)}\right\Vert _{2}} \right)^{1/M}.
\end{equation}
For a set of graphs $\{\mathcal{G}\}_{[\alpha, \beta]}$, the worst-case asymptotic convergence rate is defined as
\begin{equation}\label{gammastep}
\gamma_{asym}:= \mathop {\sup }\limits_{\{\mathcal{G}\}_{[\alpha, \beta]}} \rho_{asym}.
\end{equation}
It is easy to verify that $\rho_{asym}=\underset{M\rightarrow \infty }{\lim } {\rho_M}^{1/M}$ and $\gamma_{asym}=\underset{M\rightarrow \infty }{\lim } {\gamma_M}^{1/M}$, where $\rho_M$ is defined in (\ref{convrate}) and $\gamma_M$ is defined in (\ref{gamma}).

\begin{theorem}
For the MAS (\ref{agent}) on the set of connected graphs $\{\mathcal{G}\}_{[\alpha, \beta]}$ and under the control of a general time-varying protocol (\ref{protocol}), we have
\begin{equation}
\min_{\{\varepsilon (k)\}}{\gamma_{asym}}
= \frac{\sqrt {\beta /\alpha }-1}{\sqrt {\beta /\alpha }+1}.
\end{equation}
\end{theorem}

\begin{IEEEproof}
It is obvious that
\begin{equation}
\left( \frac{\sqrt {\beta /\alpha }+1}{\sqrt {\beta /\alpha }-1}\right)^M \geq
\left\vert {{g_M}(0)} \right\vert \geq
\frac{1}{2} \left( \frac{\sqrt {\beta /\alpha }+1}{\sqrt {\beta /\alpha }-1}\right)^M .
\end{equation}
Then
\begin{equation}
\frac{\sqrt {\beta /\alpha }+1}{\sqrt {\beta /\alpha }-1} \geq
\underset{M\rightarrow \infty }{\lim }{\left\vert {{g_M}(0)} \right\vert}^{\frac{1}{M}} \geq
\underset{M\rightarrow \infty }{\lim }{\left(\frac{1}{2}\right)^{\frac{1}{M}}} \frac{\sqrt {\beta /\alpha }+1}{\sqrt {\beta /\alpha }-1}
=\frac{\sqrt {\beta /\alpha }+1}{\sqrt {\beta /\alpha }-1}.
\end{equation}
Therefore
\begin{equation}
\underset{M\rightarrow \infty }{\lim }{\left\vert {{g_M}(0)} \right\vert^{\frac{1}{M}}}
=\frac{\sqrt {\beta /\alpha }+1}{\sqrt {\beta /\alpha }-1}.
\end{equation}
Let ${\gamma^*_{asym}}$ be the optimal value of $\min_{\{\varepsilon (k)\}}{\gamma_{asym}}$. Then
\begin{equation}
{\gamma^*_{asym}}=\underset{M\rightarrow \infty }{\lim }{\left({\gamma^*_{M}}\right)^{\frac{1}{M}}}
=\underset{M\rightarrow \infty }{\lim }{\left\vert \frac{1}{{g_M}(0)} \right\vert^{\frac{1}{M}}}
=\frac{\sqrt {\beta /\alpha }-1}{\sqrt {\beta /\alpha }+1}.
\end{equation}
\end{IEEEproof}

\section{Illustrative Examples}

In this section, we evaluate our methods by simulation and numerical experiments. Experiments have been designed to study the behaviour of the graph filters (\ref{eq23}) in Theorem 3 and (\ref{chebyh}) in Theorem 4, respectively.
We also evaluate the convergence rates on different networks and compare the two proposed methods with the best constant control gain proposed in \cite{[16]}.
We use the evaluation results to draw the guidelines for the proper choices of the design method and design parameters of graph filters, irrespective of the network structures.

\subsection{Worst-case convergence rate}

\begin{figure}
\setcounter{subfigure}{0}
\centering
\subfigure[The designed filters with $M=2$]{
\includegraphics[scale=0.32]{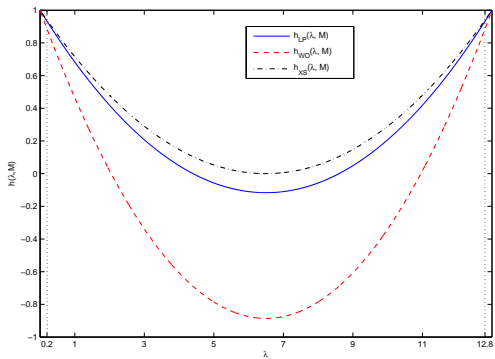}
\label{fig7a}}
\subfigure[The designed filters with $M=3$]{
\includegraphics[scale=0.32]{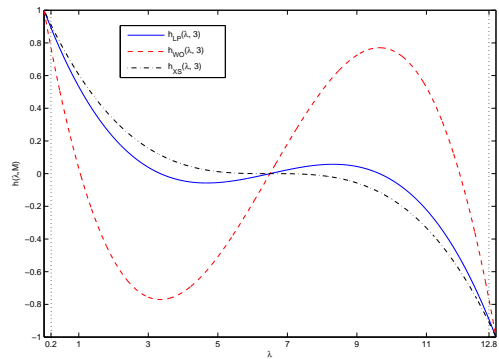}
\label{fig7b}}\\
\subfigure[The designed filters with $M=4$]{
\includegraphics[scale=0.32]{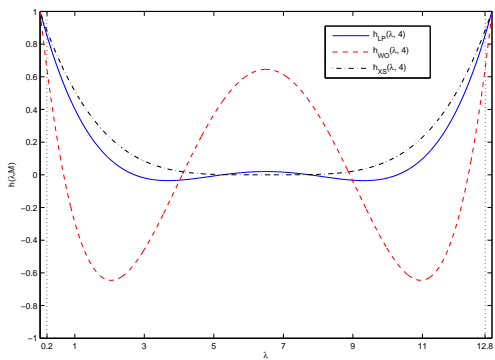}
\label{fig7c}}
\subfigure[The designed filters with $M=5$]{
\includegraphics[scale=0.32]{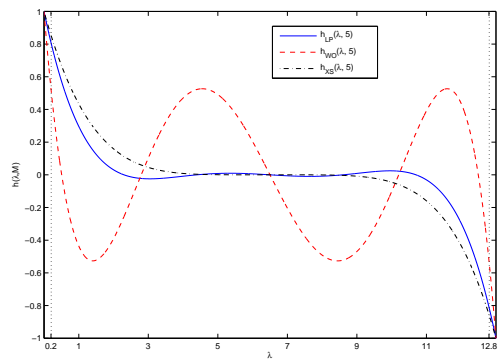}
\label{fig7d}}
\caption{The frequency responses of graph filters designed by the listed methods.}
\label{fig7}
\end{figure}

We start with MASs on connected graphs in the set $\{\mathcal{G}\}_{[0.2, 12.8]}$.
According to Theorems 3 and 4, the graph filters designed by Lagrangian polynomial (LP) interpolation method (\ref{Equ_gain}) and the worst-case optimal (WO) interpolation method (\ref{chebyh}) are respectively
\begin{eqnarray}
h_{LP}(\lambda, M) = \prod\limits_{k=1}^{M} {\left(1 - \frac{1}{0.2+\frac{12.6}{M+1}k}\lambda \right)}\label{LP}
\end{eqnarray}
and
\begin{eqnarray}
h_{WO}(\lambda, M) = \prod\limits_{k=1}^{M} {\left(1 - \frac{1}{{6.3}\cos \frac{{2k
- 1}}{{2M}}\pi + {6.5}}\lambda \right)}.\label{WO}
\end{eqnarray}
It is proved in \cite{[16]} that the best constant control gain is $\varepsilon(k) \equiv {\varepsilon_c} = \frac{2}{{0.2 + 12.8}} = \frac{1}{6.5}$, which can be regard as a first-order graph filter
$h_{XS}(\lambda, 1) = {\left(1-\frac{\lambda}{6.5} \right)}$ giving
\begin{equation}
h_{XS}(\lambda, M) = {\left(1-\frac{\lambda}{6.5} \right)^M}.\label{XS}
\end{equation}
\begin{table}
\caption{\label{rho}Worst-case convergence rate $\gamma_M$ of three methods with different $M$ ($\{\mathcal{G}\}_{[0.2, 12.8]}$)}
\centering
\begin{tabular}{c|c|c|c|c}
\hline
$M$ & $M=2$ & $M=3$ & $M=4$ & $M=5$ \\
\hline
\hline
\makecell[bc]{Lagrange polynomial interpolation method (\ref{LP})} \\ $h_{LP}(\lambda, M) = \prod\limits_{k=1}^{M} {\left(1 - \frac{1}{0.2+\frac{12.6}{M+1}k}\lambda \right)}$ & $0.9324$ & $0.8925$ & $0.8513$ & $0.8097$ \\
\hline
\makecell[bc]{Worst-case optimal interpolation method (\ref{WO})} \\ $h_{WO}(\lambda, M) = \prod\limits_{k=1}^{M} {\left(1 - \frac{1}{{6.3}\cos \frac{{2k- 1}}{{2M}}\pi + {6.5}}\lambda \right)}$ &
$0.8858$ & $0.7706$ & $0.6456$ & $0.5268$ \\
\hline
\makecell[bc]{Fast linear iterations approach (\ref{XS})} \\ $h_{XS}(\lambda, M) = {\left(1-\lambda/6.5 \right)^M}$ & $0.9394$ & $0.9105$ & $0.8824$ & $0.8554$ \\
\hline
\end{tabular}
\end{table}

\subsection{Exact convergence rate on some specific graphs}

\emph{1) The star graph $\mathcal{G}_1$:} The nonzero eigenvalues of the graph Laplacian $\mathcal{L}_{\mathcal{G}_1}$ are
$\left\{\lambda_i\right\}_{\mathcal{G}_1}=\left\{1,12\right\}.$
The exact convergence rates of MAS on $\mathcal{G}_1$, calculated by (\ref{rhoM}).
It can be seen that as $M$ increases, the exact convergence rates resulting from the Lagrangian polynomial interpolation method (\ref{eq23}) and the best constant
control gain in \cite{[16]} both monotonically decrease to zero, but that of the worst-case optimal interpolation method (\ref{chebyh}) oscillating to zero.
It appears that the Lagrangian polynomial interpolation method is always better than
the fast linear iterations approach,
but the worst-case optimal interpolation method may be worse than the latter, or even the worst of all the three methods when $M=5$.

The exact convergence rate $\rho_M$ resulting from the worst-case optimal interpolation method drops very fast when $M=3$ because the corresponding roots of the graph filter ${{h_{WO}}(\lambda, 3)}$ are $\{r_k\}=\{1.044, 6.5, 11.956\}$,
which are close to the nonzero eigenvalues $\{\lambda_i\}_{\mathcal{G}_1}$.
Thus, for the star graph $\mathcal{G}_1$, there is no need to increase $M$ to accelerate the convergence rate. One only needs to select the $3$-periodic control sequence $\varepsilon(k)=\{\frac{1}{11.956}, \frac{1}{6.5}, \frac{1}{1.044}\}$ to achieve fast consensus.

\emph{2) The cycle graph $\mathcal{G}_2$:} The nonzero eigenvalues of the graph Laplacian matrix $\mathcal{L}_{\mathcal{G}_2}$ are
$\left\{\lambda_i\right\}_{\mathcal{G}_2}=\{0.2679, 1, 2, 3, 3.7321, 4\}.$
The exact convergence rates of MAS on $\mathcal{G}_2$ calculated by (\ref{rhoM}).
As the eigenvalues are distributed uniformly on $[0, 4]$ and $\lambda_2=0.2679$ which is
close to $\alpha=0.2$, the curves of $\rho_M$ for the three methods are similar
to those of the worst case convergence rate $\gamma_M$, and the worst-case
polynomial interpolation method gives the fastest convergence rate.

\emph{3) The graph $\mathcal{G}_3$ generated by a small-world network model:} We randomly generated $50$ small-world networks with $12$ agents, and choose a rare case $\mathcal{G}_3$ to demonstrate that the worst-case optimal interpolation method may not show its advantages for specific graphs. The nonzero eigenvalues of the graph Laplacian matrix $\mathcal{L}_{\mathcal{G}_3}$ are \[\left\{\lambda_i\right\}_{\mathcal{G}_3}=\{0.655, 1.2694, 1.9964, 2.8578, 3.6319, 3.8860, 5.0364, 5.2884, 5.7759, 6.4118, 7.1909\}.\]
The exact convergence rates of MAS on $\mathcal{G}_3$, calculated
by (\ref{rhoM}.
It can be seen that the control sequence designed by the Lagrange polynomial interpolation method has the
fastest convergence rate, and for $M \leq 5$, the control sequence designed by the worst-case optimal
interpolation method has the slowest convergence rate.
This shows that for a given graph, the convergence rates resulting from interpolation methods are closely related to the spectral distribution of the graph Laplacian matrix.

\emph{4) The path grpah $\mathcal{G}_4$:} The nonzero eigenvalues of the graph Laplacian matrix $\mathcal{L}_{\mathcal{G}_4}$ are
\[\left\{\lambda_i\right\}_{\mathcal{G}_4}=\{0.2679, 1, 2, 3, 3.7321\}.\]
It can be seen that $\left\{\lambda_i\right\}_{\mathcal{G}_4}=\left\{\lambda_i\right\}_{\mathcal{G}_2} \backslash \{4\}$.
This means that although the network structure and the number of agent are different, the Laplacian spectral distributions of the graph $\mathcal{G}_4$ and the graph $\mathcal{G}_2$ are nearly the same.
Hence, the exact convergence rates of MAS on $\mathcal{G}_4$ calculated
by (\ref{rhoM}) are similar to those on $\mathcal{G}_2.$
This shows that for different graphs with similar Laplacian spectral distributions, the convergence performance of the three methods is almost the same.

\begin{table}
\caption{\label{rhoGcompar}The exact convergence rate $\rho_M$ of three methods on four specific graphs }
\centering
\begin{tabular}{c|c|c|c|c|c|c|c|c|c|c|c|c}
\hline
\multirow{2}{*}{Method} & \multicolumn{3}{c|}{Star graph ${\mathcal{G}_1}$} & \multicolumn{3}{c|}{Cycle graph ${\mathcal{G}_2}$} & \multicolumn{3}{c|}{Small-World graph ${\mathcal{G}_3}$} & \multicolumn{3}{c}{Path graph ${\mathcal{G}_4}$} \cr \cline{2-13}
& LP & WO & XS & LP & WO & XS & LP & WO & XS & LP & WO & XS \cr
\hline
\hline
$M=2$ & 0.6829 & 0.4645 & 0.7160 & 0.9099 & 0.8478 & 0.9193 & 0.7264 & 0.8807 & 0.7549 & 0.9099 & 0.8478 & 0.9193
\cr \hline
$M=3$ & 0.5321 & 0.0328 & 0.6059 & 0.8577 & 0.7556 & 0.8814 & 0.5906 & 0.7556 & 0.6559 & 0.8577 & 0.7556 & 0.8814
\cr \hline
$M=4$ & 0.4024 & 0.2907 & 0.5127 & 0.8044 & 0.6449 & 0.8451 & 0.4693 & 0.6454 & 0.5699 & 0.8044 & 0.6449 & 0.8451
\cr \hline
$M=5$ & 0.2961 & 0.4363 & 0.4338 & 0.7515 & 0.4696 & 0.8103 & 0.3656 & 0.5231 & 0.4952 & 0.7515 & 0.4362 & 0.8103
\cr \hline
\end{tabular}
\end{table}

Table \ref{rhoGcompar} gives the exact values of the convergence rate $\rho_M$ on the four specific graphs fot the
control period $M=2, 3, 4, 5$. Compared to the best constant control gain proposed in \cite{[16]}, the Lagrange
polynomial interpolation method always has a faster convergence rate, which corroborates the analysis in Remark \ref{Remark7}.
\begin{remark}
Although none of the three methods can attain the fastest convergence rate for all the graphs in $\mathcal{G}_{[0.2, 12.8]}$, the best control sequence for average consensus is always designed by one of the methods we proposed.
\end{remark}


\subsection{Consensus performance}

To compare the evolution of the MAS states of the three methods on the four specific graphs, we take the control period $M=3$ as an example. Then the graph filters (\ref{LP})-(\ref{XS}) become
\begin{eqnarray*}
h_{LP}(\lambda, 3) &=& \left(1 - \frac{\lambda}{3.35} \right)\left(1 - \frac{\lambda}{6.5} \right)\left(1 - \frac{\lambda}{9.65} \right), \\
h_{WO}(\lambda, 3) &=& \left(1 - \frac{\lambda}{1.044} \right)\left(1 - \frac{\lambda}{6.5} \right)\left(1 - \frac{\lambda}{11.956} \right), \\
h_{XS}(\lambda, 3) &=& {\left( 1 - \frac{\lambda}{6.5} \right)}^3.
\end{eqnarray*}

For the initial states of the agents randomly taken from the interval $[0, 10]$, the MASs on the four specific graphs reaches average consensus asymptotically under the control protocols of the above graph filters.

For the star graph $\mathcal{G}_1$, the MAS under the control protocol of the worst-case optimal interpolation method converges very fast, whereas
for the cycle graph $\mathcal{G}_2$ and the path graph $\mathcal{G}_3$, the convergence rates are much slower due to the distributions of Laplacian spectra.
For the graph $\mathcal{G}_4$, the consensus performance of the worst-case optimal interpolation method is the worst of all the three methods.

\subsection{Consensus on Large-scale networks}

\begin{figure}
  \centering
  \centerline{\includegraphics[scale=0.5]{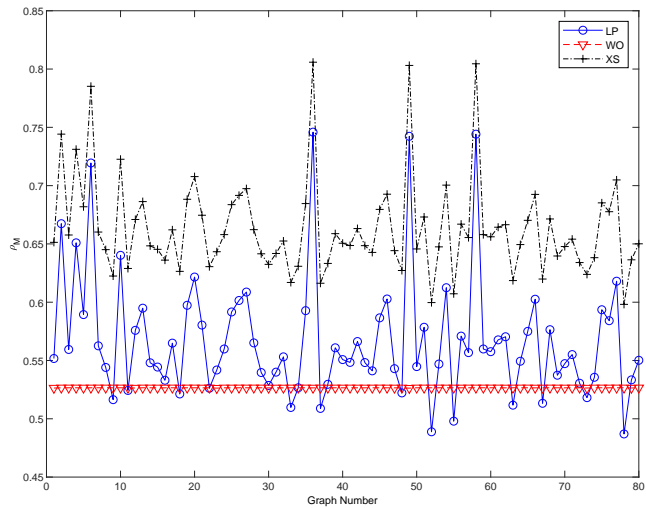}}
  \caption{The exact convergence rate $\rho_M$ for $80$ random graphs by different methods with $M=5$.}
  \label{fig14}
\end{figure}

We generate $80$ random connected graphs of $100$ nodes in the set $\{\mathcal{G}\}_{[0.2, 12.8]}$. To verify the effectiveness of the graph filters shown above, we plot in Fig. \ref{fig14} the exact convergence rates $\rho_{M}$ resulting from these filters. It can be seen that the exact convergence rate $\rho_M$ of the Lagrange polynomial interpolation method is always smaller than that of the fast linear iteration approach. This indicates that the periodic control sequence from the Lagrange polynomial interpolation method has better consensus performance compared with the constant control gain proposed in \cite{[16]}.
For the worst-case optimal interpolation method, the exact convergence rate $\rho_{WO}$ is almost a straight line around the worst-case convergence rate $\gamma_{WO}=0.5268$, with very little fluctuation.
It is noted that in about $15\%$ of the $80$ graphs, the exact values of $\rho_M$ for the worst-case optimal interpolation method are larger than those of the Lagrange polynomial interpolation method.
This shows that the worst-case optimal approximation method is optimal in the worst case, but not always good for some specific examples.

\section{Conclusion}

This paper has established the explicit connection between MAS consensus and spectral filtering of graph signals. Based on this connection, we have developed an effective new approach to the analysis of MAS consensus and the design of effective control protocols in the graph spectrum domain. Using this approach, we have obtained a number of new analysis results and effective new design methods, as summarized in Abstract, for the consensus of MASs on the uncertain graphs. We have presented numerical examples to show the validity, effectiveness and advantages of these results and methods. All of these have demonstrate that this new graph spectrum domain approach can overcome the difficulties of the existing time-domain methods in dealing with uncertain networks, and allows us to obtain more precise analysis results and deeper insights, more effective design methods and better consensus performance than those of the existing works on MAS consensus.

For simplicity and also due to space limit, we have presented our results only for the first order deterministic MAS on undirected graphs. In fact, all the results presented in this paper can be extended to the high-order and stochastic MASs on the directed graphs. These results will be reported elsewhere.

\end{document}